# Retrosynthesis of Synthetic Media for Explainable AI Provenance Forensics

Yijie Lin, Ching-Chun Chang, *Senior Member, IEEE*, Isao Echizen, *Senior Member, IEEE*, Hui Li, *Senior Member, IEEE*, and Chin-Chen Chang, *Fellow, IEEE*

***Abstract*—With the rapid proliferation of generative models on Machine Learning as a Service (MLaaS) platforms, reliably tracing the provenance of synthetic media without modifying generator architectures or parameters remains a major challenge. In this work, we propose a self-referential retrosynthesis framework for explainable AI provenance forensics under a fixed-generator setting. The framework leverages a jointly optimized encoder-decoder pair to implement a self-embedding mechanism that enables round-trip consistency verification. During inference, client inputs are first encoded and then processed by the generator to produce outputs with high visual fidelity. For forensic verification, the consistency between the resynthesized image and the query image is analyzed to determine whether the image originates from the target generative model. Our approach eliminates the need for watermark embedding or modifications to the generation process. Experimental results show that images generated from encoded inputs maintain visual quality comparable to original generator outputs, while decoded images reliably trace back to their corresponding source inputs. Furthermore, the framework provides interpretable evidence for generative content provenance, establishing a practical tool for explainable generative AI forensics.**

## I. INTRODUCTION

In recent years, generative artificial intelligence (AI) has made significant progress and has been widely deployed on Machine Learning as a Service (MLaaS) platforms [1-3]. Through standardized application programming interfaces (APIs), clients can conveniently submit input data and obtain high-quality generated results. Despite its practicality and accessibility, this service paradigm also raises pressing challenges in privacy and intellectual property protection [4-5], such as membership inference attacks, where an adversary can infer whether certain data were included in the training set [6], and ambiguity attacks, which make model ownership claims harder to verify [7]. Among various security challenges, provenance verification remains particularly difficult, as it is often non-trivial to establish a reliable link between a generated output and the source input that gave rise to it. If such a reliable link can be established, it enables several practical capabilities that are essential for generative provenance forensics:

- Traceability: The originating input can be recovered from the generated content, mitigating input ownership ambiguity. For example, when a client submits proprietary inputs (e.g., design drafts, personal photographs, or sketches) to obtain generated results, traceability allows the client to demonstrate that a disputed output is indeed derived from their input.
- Accountability: The generated content can be verified against a claimed generator, supporting responsibility attribution when multiple platforms may provide similar generation services. This is particularly important in MLaaS workflows where synthetic images may be redistributed without reliable provenance metadata.
- Explainability: The linkage between the recovered input and the generated output can be supported by an interpretable and auditable chain of evidence, rather than an opaque decision. Such explainable evidence strengthens forensic reliability by enabling stakeholders to understand why a particular input-output relationship is claimed, especially when a creator challenges the origin of circulated content.

Modern generative models have evolved through three major paradigms, namely variational autoencoders (VAE) [8], generative adversarial networks (GAN) [9], and diffusion models [10], upon which numerous image generation systems have been developed [11-14]. As these models become increasingly integrated into real-world production pipelines, disputes and ambiguities may arise in determining whether a given image was produced by a claimed generator and which input should be regarded as its legitimate origin. Such uncertainties hinder responsibility attribution and weaken the evidential basis required for audit and governance, particularly when generated content is disseminated beyond its original generation context.

Recent industry efforts, such as the Coalition for Content

This work was supported in part by JSPS KAKENHI under Grant JP21H04907 and Grant JP24H00732, in part by JST CREST under Grant JPMJCR20D3 and Grant JPMJCR2562 including AIP challenge program, in part by JST AIP Acceleration under Grant JPMJCR24U3, and in part by JST K Program under Grant JPMJKP24C2, Japan. *(Corresponding author: Ching-Chun Chang.)*

Y. Lin and C.-C. Chang (Chin-Chen Chang) are with the Department of Information Engineering and Computer Science, Feng Chia University, Taichung 407, Taiwan (emails: p1263670@o365.fcu.edu.tw; ccc@o365.fcu.edu.tw).

C.-C. Chang (Ching-Chun Chang) and I. Echizen are with the Information and Society Research Division, National Institute of Informatics, Tokyo 101-8430, Japan (emails: ccchang@nii.ac.jp; iechizen@nii.ac.jp).

I. Echizen is also with the Graduate School of Information Science and Technology, University of Tokyo, Tokyo 113-8656, Japan.

H. Li is with the Shenzhen Graduate School, Peking University, Shenzhen 518055, China (email: lih64@pkusz.edu.cn).

Provenance and Authenticity (C2PA), focus on standardizing content provenance through cryptographically signed metadata and manifests attached to media assets. While such metadata-centric frameworks provide an important foundation for content attribution, they operate outside the generative process and rely on the integrity and availability of external metadata. As a result, provenance information may be lost, stripped, or invalidated during common content redistribution and transformation workflows, and these approaches do not directly establish an intrinsic linkage between a generated output and the underlying generative model or source input.

In contrast to metadata-centric frameworks, existing approaches related to generative content provenance can be broadly categorized into three types. The first type embeds watermarks or feature markers during the generation process to facilitate subsequent identification [15-22]. Notable industrial deployments include SynthID developed by Google. While effective in signaling ownership, these methods operate exclusively on the final outputs and fundamentally lack the capability to trace the original source inputs driving the generative process. The second type relies on external discriminators or detection networks for post hoc classification of generated images [23-27]. Although such methods do not alter the generator, they typically provide limited traceability to the originating input and may exhibit reduced robustness under distribution shifts. The third type performs source tracing through model fingerprints or model-specific characteristics [28-31], enabling fine-grained attribution in certain settings but often assuming access to model-dependent information.

Motivated by these limitations, this paper proposes a self-referential retrosynthesis framework for generative provenance forensics, integrating self-embedding, retrosynthesis, and round-trip provenance forensics. The key idea is to induce intrinsic traceability in the input space through self-embedding, enabling the originating input to be recovered from generated content via retrosynthesis. Building upon this recoverability, we further introduce a round-trip forensic protocol that validates the recovered input by re-synthesis under the claimed generator, producing an interpretable and auditable chain of evidence for authentication and attribution. Unlike prior studies that primarily emphasize model-centric ownership protection, our framework explicitly addresses the often-overlooked problem of input ownership in generative services, ensuring that both source inputs and generated outputs can be systematically traced and verified.

Fig. 1 illustrates the application scenario of the proposed self-referential retrosynthesis framework, which supports generative provenance forensics on an MLaaS platform. During the generation phase, a client submits a source image that is processed via the server API by the deployed generative AI, and the resulting synthetic image is returned to the client. The framework also provides a provenance verification phase, allowing a query image to be submitted to the server API. The server returns a retrosynthetic image along with a verdict on whether the image was generated by the deployed platform.

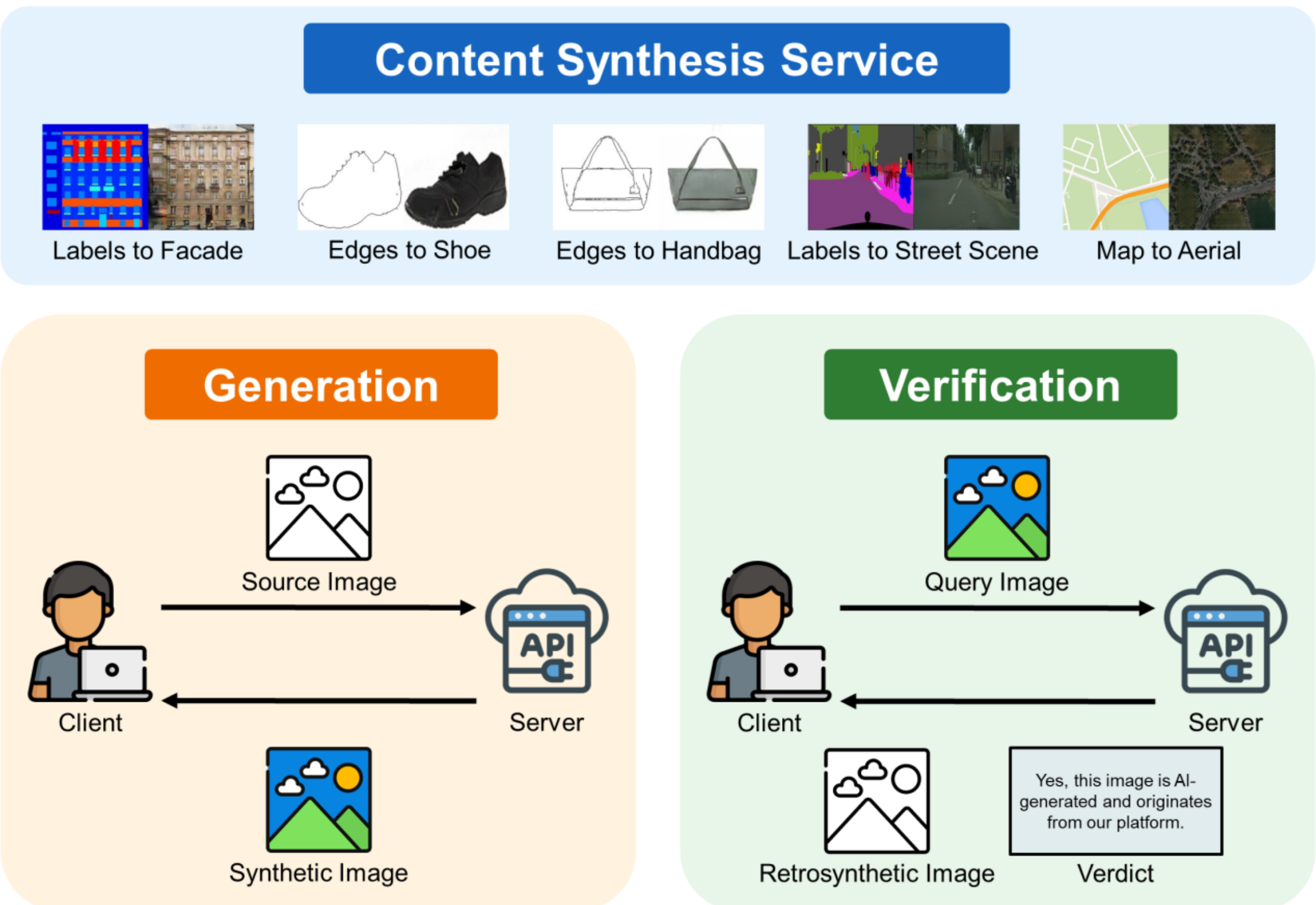

**Fig. 1.** Application scenario of the proposed self-referential retrosynthesis framework for generative provenance forensics.

The main contributions of this paper are summarized as follows:

- **Self-Embedding:** We introduce a self-embedding mechanism to induce intrinsic traceability in generative inputs, allowing provenance cues to be carried through the generation process.
- **Retrosynthesis:** We formulate retrosynthesis as a principled recovery process that maps generated content back to its originating input, offering an interpretable and back-traceable provenance path.
- **Round-Trip Provenance Forensics:** We propose a round-trip forensic protocol that validates recovered inputs by re-synthesis under the claimed generator, producing an explainable chain of evidence for authentication and attribution.

The remainder of this paper is organized as follows. Section II reviews the related literature on provenance tracing. Section III presents the proposed framework, including the problem formulation, network architectures, learning objectives, and the procedures for training, inference, and verification. Section IV describes the experimental setup and reports the evaluation results, covering encoding-decoding fidelity, generation stability, round-trip attribution, and generalization across models and tasks. Finally, Section V concludes the paper.

## II. Related Works

This section reviews prior studies that are most relevant to traceable and verifiable generative AI. We first summarize representative watermarking approaches that embed hidden ownership signals into generated content. We then discuss detection methods that identify AI-generated content using external classifiers, while generally lacking the ability to trace the source model. Finally, we review fingerprinting techniques that aim to attribute generated content to specific models or clients, often assuming access to model internals or training data.

### *A. Watermarking for Generative Artificial Intelligence*

Generative AI watermarking techniques aim to embed hidden and verifiable watermarks into model-generated content to assert ownership. Zhang et al. [18] introduced a task-independent barrier appended to the target model, embedding a uniform and invisible watermark into its output. Qin et al. [19] designed a deep noise simulation network to model the fusion process of real noise, producing highly robust watermarked images. Huang et al. [20] enhanced image features in inconspicuous and textured regions to embed watermarks, guided by attention masks. Yuan et al. [21] injected watermarks into a stable diffusion model, generating watermarks based on predefined cues. Xu et al. proposed InvisMark [22], specifically for high-resolution AI-generated images, embedding imperceptible and highly robust watermarks. These techniques rely entirely on post-generation modifications and cannot trace the originating source input.

### *B. Detection for Generative Artificial Intelligence*

Generative AI detection technologies focus on determining whether content is generated by AI models, to support content moderation and platform governance. In 2020, Wang et al. proposed CNNSpot [23], which distinguishes real images from those generated by convolutional neural networks. In 2023, Ojha et al. proposed UniversalFakeDetect (UFD) [24], using a feature space not explicitly trained to distinguish between real and fake images for real-fake classification. Also in 2023, Wang et al. proposed Diffusion Reconstruction Error (DIRE) [25], which measures the error between an input image and its reconstructed image from a pre-trained diffusion model, thus distinguishing between real images and diffusion-generated images. In 2024, Tan et al. proposed Neighboring Pixel Relationships (NPR) [26], capturing and representing general structural artifacts generated by upsampling operations to achieve detection of generalizable deepfakes. In 2025, Zhang et al. proposed Variational Information Bottleneck (VIB-Net) [27], utilizing variational information bottlenecks to enhance feature learning relevant to authentication tasks, improving detection accuracy. These technologies are effective for identifying AI-generated content while relying on external classifiers and not supporting source tracing.

### *C. Fingerprinting for Generative Artificial Intelligence*

Generative AI fingerprinting technologies aim to assign distinct identifiers to content from different models, versions, or clients, enabling source tracing and accountability. Marra et al. [28] demonstrated in 2019 that each GAN leaves a unique fingerprint in the images it generates. In the same year, Yu et al. proposed GAN Fingerprints (GAN-F) [29], systematically investigating how to learn model-specific fingerprints for image source tracing and for distinguishing real images from GAN-generated images. Subsequently, Yu et al. [30] introduced artificial fingerprints into training data in 2021, allowing such fingerprints to be transferred to the trained generative models and to manifest in the resulting deepfake images. In 2023, Zhu et al. [31] proposed a constrained fingerprint generation strategy within the GAN framework, ensuring that the enhanced fingerprints remain consistent with real fingerprint characteristics. These technologies enable fine-grained source identification while often assuming access to model internals or training data.

## III. Proposed Methodology

This section presents our methodology for generative provenance forensics. We first formulate the problem and describe the relevant application scenarios. We then detail the proposed framework in terms of network architectures, optimization objectives, and end-to-end procedures for training, inference, and round-trip verification.

### *A. Problem Formulation*

The proposed framework addresses three key issues. First, it preserves the original functionality of the MLaaS platform, ensuring that the quality of generative outputs is not affected.

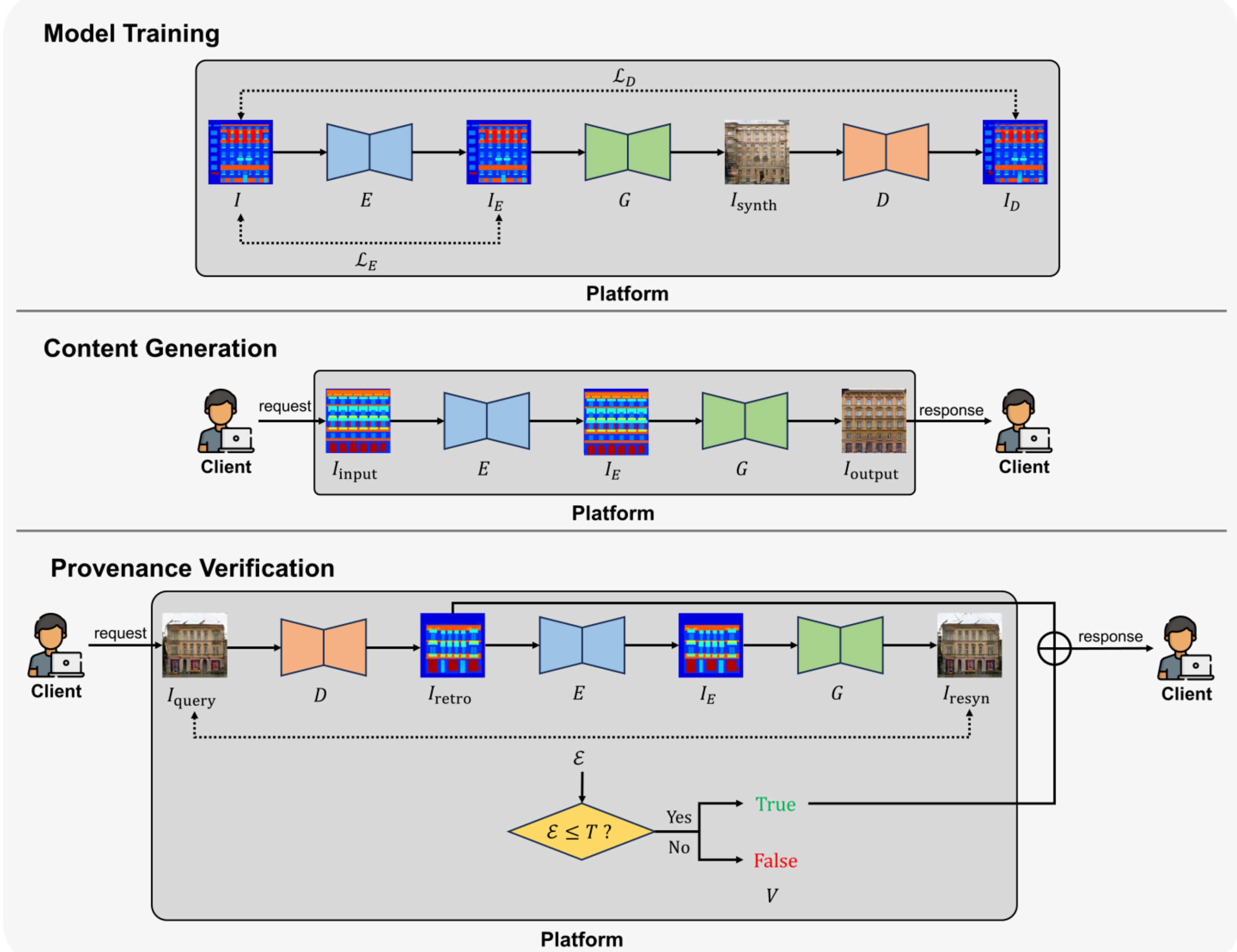

**Fig. 2.** Overview of the proposed self-referential retrosynthesis framework for generative provenance forensics.

Second, it guarantees that the ownership of client inputs is maintained by enabling traceability of the source image. Third, it enables verification of a query image to determine whether it was generated by the MLaaS platform, allowing reliable distinction from both real images and images generated by other generative AI models.

### *B. Network Architectures*

The proposed self-referential retrosynthesis framework for generative provenance forensics comprises an encoder, a fixed generator, and a decoder, as shown in Fig. 2.

- Encoder ($E$): The encoder maps the input source image to an encoded image. It employs a U-Net [32] structure with multi-scale feature fusion to ensure that the encoded image preserves the original content while embedding traceable signals.
- Generator ($G$): The generator is the originally deployed model, fixed and directly invoked without any modification, producing outputs nearly identical to those generated from the original inputs.
- Decoder ($D$): The decoder receives the output from the generator and traces back the source image. It also employs a U-Net architecture to ensure high-fidelity retrosynthetic images.

This architecture enables traceable information to be embedded into the generation process, ensuring that the source image can be reliably traced through the decoder.

### *C. Optimization Objectives*

The proposed architecture jointly optimizes the encoder and the decoder using a unified loss function. To ensure that the encoded image $I_E$ retains information from the source image $I$, the encoder loss is defined as

$$\mathcal{L}_E = ||I_E - I||_1 \ . \tag{1}$$

To ensure that the decoded image $I_D$ can be traced back to the source image $I$, the decoder loss is defined as

$$\mathcal{L}_D = ||I_D - I||_1 \ . \tag{2}$$

The overall loss function is defined as

$$\mathcal{L}_{\text{total}} = \lambda_E \mathcal{L}_E + \lambda_D \mathcal{L}_D \ . \tag{3}$$

Here, $\lambda_E$ and $\lambda_D$ are weighting hyperparameters used to balance the encoder and decoder losses. The Adam optimizer

[33] is employed to achieve stable convergence during training.

The model objective is to minimize the total loss with respect to the encoder and decoder, that is, $\min_{E,D}\mathcal{L}_{\text{total}}$.

*D. Model Training Procedures*

The model training phase aims to enable the encoder and decoder to collaboratively learn traceable information from the source image through self-embedding. As shown in Fig. 2, the source image $I$ is first fed into the encoder, which produces an encoded image $I_E$, denoted as

$$I_E = E(I)\ . \tag{4}$$

The encoded image $I_E$ is then input to the generator, which produces a synthetic image $I_{\text{synth}}$, denoted as

$$I_{\text{synth}} = G(I_E)\ . \tag{5}$$

Subsequently, the synthetic image $I_{\text{synth}}$ is fed into the decoder to obtain a decoded image $I_D$, denoted as

$$I_D = D(I_{\text{synth}})\ . \tag{6}$$

During training, the encoder and decoder parameters are jointly optimized iteratively until convergence, with the generator remaining fixed.

Algorithm 1 details the training procedure of the proposed framework, which jointly optimizes the encoder and decoder to ensure input recoverability.

**Algorithm 1 Model Training**

**Input:** Training dataset $TD$; Hyperparameters $\lambda_E$, $\lambda_D$

**Output:** Optimized encoder $E$; Optimized decoder $D$

1: **Initialize** $E$ and $D$ with random weights
2: **Load generator** $G$ (Fixed)
3: **for each** batch of images $I \in TD$ **do**
4: ▷ Forward Pass
5: $I_E \leftarrow E(I)$
6: $I_{\text{synth}} \leftarrow G(I_E)$
7: $I_D \leftarrow D(I_{\text{synth}})$
8: ▷ Loss Calculation
9: $\mathcal{L}_E \leftarrow ||I_E - I||_1$
10: $\mathcal{L}_D \leftarrow ||I_D - I||_1$
11: $\mathcal{L}_{\text{total}} \leftarrow \lambda_E\mathcal{L}_E + \lambda_D\mathcal{L}_D$
12: ▷ Back-propagation
13: **Update** $E$ and $D$ to minimize $\mathcal{L}_{\text{total}}$
14: **end for**
15: **return** $E$, $D$

*E. Content Generation Procedures*

The primary objective of the content generation phase is to generate a traceable output image. As illustrated in Fig. 2, during inference, the input image $I_{\text{input}}$ is first fed into the encoder to generate an encoded image $I_E$, which is then fed into the generator to produce the output image $I_{\text{output}}$, represented as

$$I_{\text{output}} = G(E(I_{\text{input}}))\ . \tag{7}$$

This process enables the client to obtain a traceable output image generated by the deployed generative AI on the MLaaS platform.

Algorithm 2 presents the content generation process, where the client's source image is encoded with traceable signals before generation.

**Algorithm 2 Content Generation**

**Input:** Source image $I_{\text{input}}$

**Output:** Traceable synthetic image $I_{\text{output}}$

1: $I_E \leftarrow E(I_{\text{input}})$
2: $I_{\text{output}} \leftarrow G(I_E)$
3: **return** $I_{\text{output}}$

*F. Provenance Verification Procedures*

The provenance verification phase has two core objectives: first, to trace the source image and ensure client ownership of it; and second, to determine the origin of the generative AI, specifically whether it was generated by the generative AI of the MLaaS platform. As shown in Fig. 2, in the provenance verification phase, the query image $I_{\text{query}}$ is first input to the decoder, producing a retrosynthetic image $I_{\text{retro}}$, denoted as

$$I_{\text{retro}} = D(I_{\text{query}})\ . \tag{8}$$

Subsequently, the round-trip consistency verification is performed, where the retrosynthetic image $I_{\text{retro}}$ is input to the encoder to generate an encoded image $I_E$. The encoded image is then input to the generator, producing a resynthesized image $I_{\text{resyn}}$, denoted as

$$I_{\text{resyn}} = G(E(I_{\text{retro}}))\ . \tag{9}$$

The mean absolute error $\mathcal{E}$ between the resynthesized image and the query image, i.e., the L1 loss, is calculated as

$$\mathcal{E} = ||I_{\text{resyn}} - I_{\text{query}}||_1\ . \tag{10}$$

Finally, the verdict $V$ is defined as

$$V = \begin{cases} \text{True}, & \text{if } \mathcal{E} \le T \\ \text{False}, & \text{otherwise} \end{cases}, \tag{11}$$

where $T$ is the threshold. The query image $I_{\text{query}}$ is determined to have been generated by the deployed MLaaS platform when $\mathcal{E} \le T$, and otherwise to have originated elsewhere. The use of round-trip consistency verification allows the framework to reliably determine image provenance without requiring an additional discriminator network.

Algorithm 3 describes the provenance verification procedure, which utilizes round-trip consistency to authenticate the origin of a query image.

**Algorithm 3 Provenance Verification**

**Input:** Query image $I_{\text{query}}$; Threshold $T$

**Output:** Retrosynthetic image $I_{\text{retro}}$; verdict $V$

1: ▷ Provenance Tracing
2: $I_{\text{retro}} \leftarrow D(I_{\text{query}})$
3: ▷ Round-Trip Consistency Check
4: $I_E \leftarrow E(I_{\text{retro}})$
5: $I_{\text{resyn}} \leftarrow G(I_E)$
6: ▷ Verification Decision
7: $\mathcal{E} \leftarrow ||I_{\text{resyn}} - I_{\text{query}}||_1$
8: **if** $\mathcal{E} \le T$ **then**
9: $V \leftarrow$ True
10: **else**
11: $V \leftarrow$ False
12: **end if**
13: **return** $I_{\text{retro}}$, $V$

## IV. Experimental Results

To provide a comprehensive evaluation of the proposed framework, we organize our experiments into following parts. We first describe the experimental setup, including the computing environment and implementation details. We then examine whether self-embedding preserves input fidelity and enables accurate retrosynthesis, analyze its impact on downstream generation stability, and assess whether round-trip consistency provides reliable evidence for generator attribution. Finally, we extend the evaluation across different generative models and tasks to validate generalizability.

### *A. Experimental Setup*

This subsection specifies the computing environment, datasets, generators, and training protocol used throughout the experiments. The experiments were conducted on a workstation equipped with an Intel® Core™ i9-10900K CPU and an NVIDIA GeForce RTX™ 3090 GPU. U-Net and its variants are commonly employed as backbones in image-to-image generative models and constitute the core structure of most mainstream generators deployed on MLaaS platforms. Accordingly, Pix2Pix [11] and CycleGAN [12] were selected as the fixed generator models for the experiments, as their generator architectures are also based on U-Net, making them suitable for evaluating the proposed self-embedding framework.

The proposed framework treats the generator as fixed, with its architecture and parameters remaining unchanged throughout training and inference, while only the encoder and decoder are optimized in an end-to-end manner. Experiments were conducted on five publicly available datasets, namely the CMP Facade Database [34], edges2shoes [35], edges2handbags [36], Cityscapes [37], and Google Maps [11], which are commonly used benchmarks for evaluating image-to-image generation performance across diverse visual domains. Unless stated otherwise, experiments were conducted on the CMP Facade dataset using Pix2Pix as the default fixed generator, while the other datasets and generators were used to evaluate the applicability of the proposed framework. For all experiments, the models were trained for 100 epochs, and the weighting coefficients of the encoder loss and decoder loss were empirically determined as $\lambda_E = 1$ and $\lambda_D = 1$, respectively. The threshold for round-trip consistency verification was set to 20 based on experimental validation, effectively distinguishing whether a given image was generated by the proposed framework.

### *B. Fidelity of Encoding and Decoding*

This subsection evaluates whether self-embedding preserves the visual fidelity of the original inputs and whether retrosynthesis can accurately recover the originating inputs. Fig. 3 provides a comparison with baseline reverse models. We trained reverse models of Pix2Pix [11] and CycleGAN [12] as baselines. The results show that, although the reverse models can approximately recover the source style, their traceability is limited, whereas the proposed decoder accurately traces back to high-quality images. Table I compares the visual quality of the source and encoded images across 100 samples from the CMP Facade Database [34]. The peak signal-to-noise ratio (PSNR) reaches a maximum of 46.59 dB, with an average of 39.21 dB. The structural similarity index (SSIM) is also close to 1, indicating high similarity between the source and encoded images. Table II presents a comparison of the visual quality of the source and decoded images over the same dataset. The average PSNR is 26.09 dB, and the average SSIM reaches 0.8552, demonstrating that the decoded images maintain traceable visual quality.

TABLE I
Visual Quality Comparison Between the Source Images and the Encoded Images

| | Maximum | Average | Minimum |
|---|---|---|---|
| PSNR | 46.59 | 39.21 | 35.50 |
| SSIM | 0.9890 | 0.9770 | 0.9655 |

TABLE II
Visual Quality Comparison Between the Source Images and the Decoded Images

| | Maximum | Average | Minimum |
|---|---|---|---|
| PSNR | 35.38 | 26.09 | 21.22 |
| SSIM | 0.9336 | 0.8552 | 0.7955 |

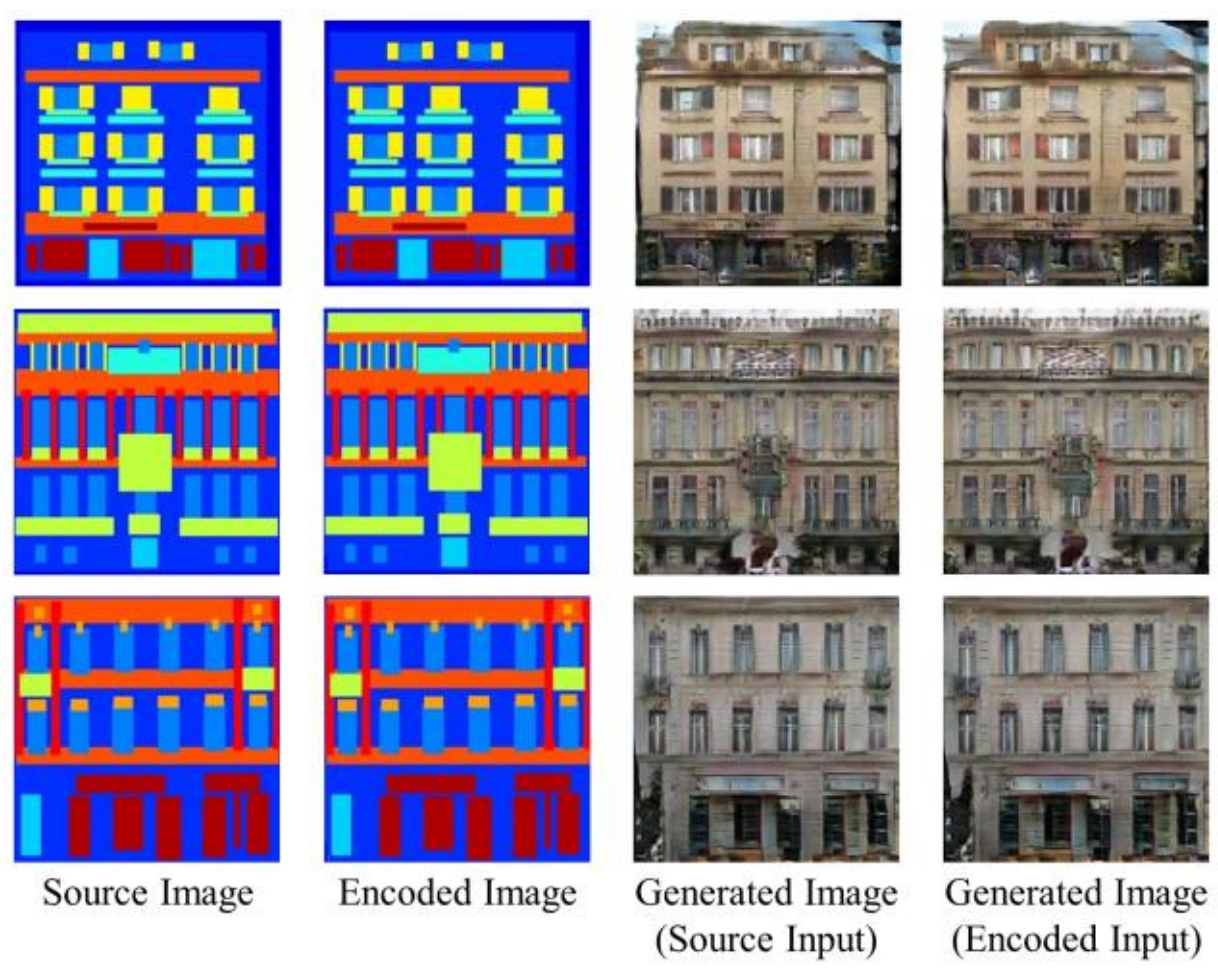

**Fig. 4.** Fidelity comparison between images generated from the source images and the encoded images.

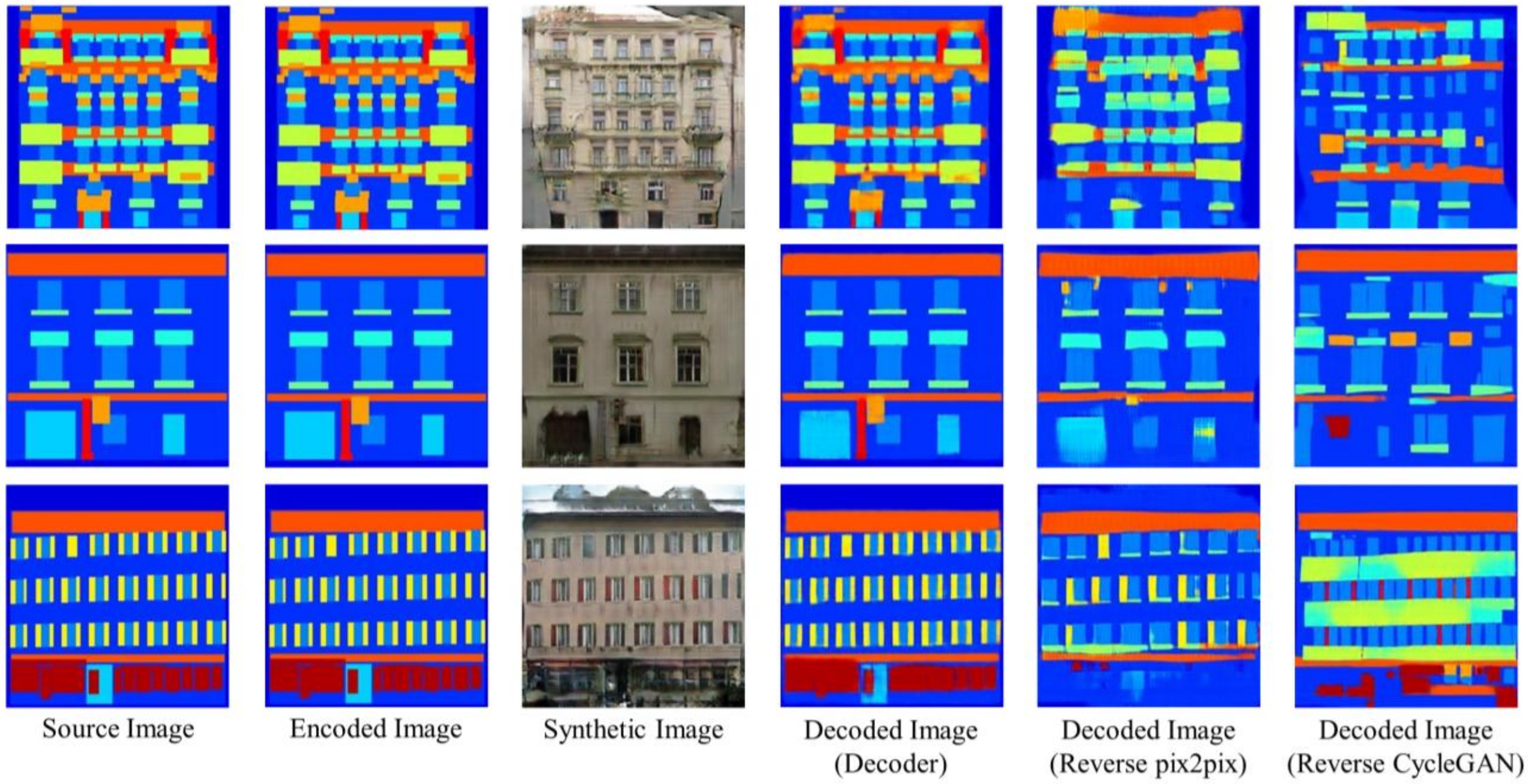

**Fig. 3.** Quality comparison with baseline reverse models.

### *C. Impact of Self-Embedding on Generation Stability*

This subsection assesses whether self-embedding alters downstream generation stability, by comparing the quality of outputs generated from embedded versus original inputs under identical settings. Fidelity is an important metric, and the proposed framework should not significantly affect the original generation task, meaning that it should preserve the original output quality. Fig. 4 presents a fidelity comparison between images generated from the source images and those generated from the encoded images. The results show that when the source and encoded images are nearly identical, the corresponding synthetic images are also highly similar. Table III quantifies the visual quality between the two types of synthetic images across 100 samples from the CMP Facade Database [34]. The average PSNR reaches 29.13 dB, and the average SSIM reaches 0.93, indicating minimal impact of the proposed framework on generation quality. Table IV further evaluates fidelity using the Learned Perceptual Image Patch Similarity (LPIPS) [38], computed with feature extraction networks AlexNet [39] and VGG [40], and the Fréchet Inception Distance (FID) [41]. The LPIPS and FID values for images generated from the source and encoded images are closely matched, demonstrating the high fidelity of the proposed framework.

TABLE III
VISUAL QUALITY COMPARISON BETWEEN IMAGES GENERATED FROM THE SOURCE IMAGES AND THE ENCODED IMAGES

| | Maximum | Average | Minimum |
|---|---|---|---|
| PSNR | 35.40 | 29.13 | 23.71 |
| SSIM | 0.9809 | 0.9317 | 0.8696 |

TABLE IV
FIDELITY EVALUATION OF IMAGES GENERATED FROM THE SOURCE IMAGES AND THE ENCODED IMAGES

| Input | LPIPS (Alex) | LPIPS (VGG) | FID |
|---|---|---|---|
| Source Images | 0.3735 | 0.4636 | 140.56 |
| Encoded Images | 0.3980 | 0.4817 | 140.70 |

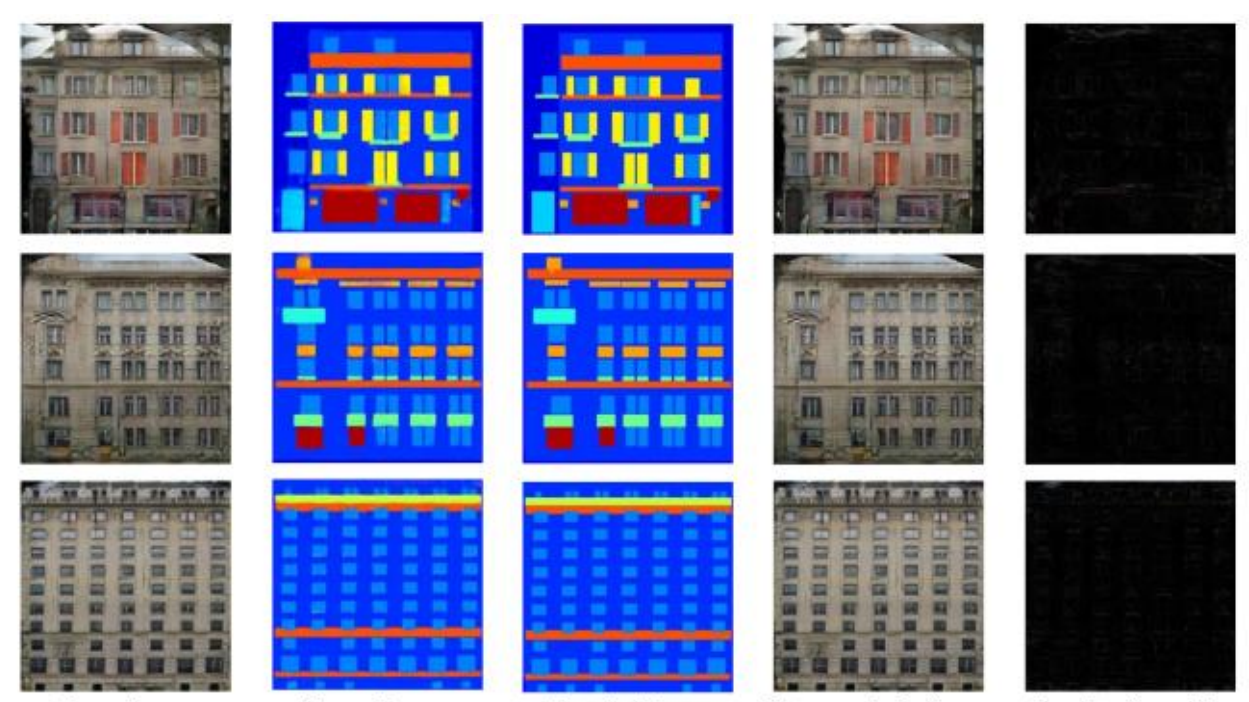

**Fig. 5.** Visualization of the provenance verification process.

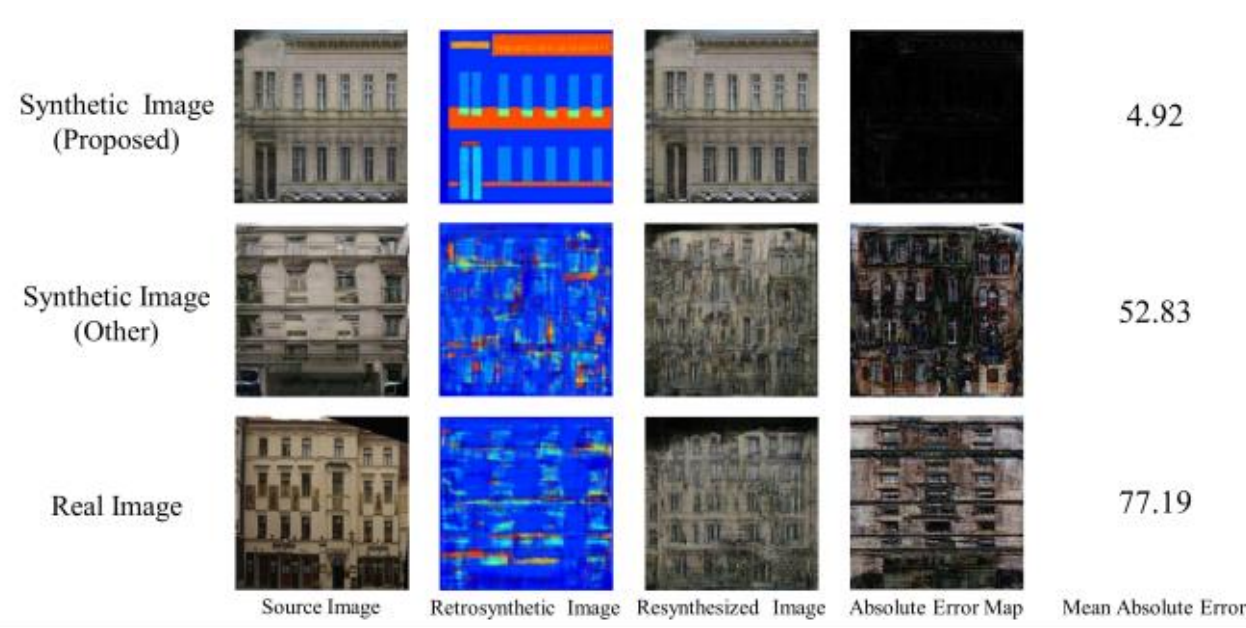

**Fig. 6.** Verification comparison among images generated by the proposed model, other generative model, and real images.

TABLE V
MEAN ABSOLUTE ERROR COMPARISON FOR DIFFERENT QUERY IMAGES

| Query Images | Maximum | Average | Minimum |
|---|---|---|---|
| Synthetic Images (Proposed) | 13.28 | 6.13 | 2.65 |
| Synthetic Images (Other) | 86.06 | 56.55 | 35.13 |
| Real Images | 87.16 | 57.42 | 33.00 |

### *D. Attributability of Generators via Round-Trip Consistency*

This subsection evaluates the extent to which round-trip consistency enables generator attribution by validating whether a given output can be reliably linked to the claimed generator through inverse recovery and forward re-synthesis. Fig. 5 illustrates the provenance verification process. When the query image is generated by the proposed framework, the resynthesized image closely matches the query image. Computing the absolute error for each pixel yields an absolute error map that is dominated by black regions, indicating minimal reconstruction error. Fig. 6 compares verification results for query images generated by the proposed framework, other generative models, and real images. When the query image is generated by the proposed framework, the source image can be accurately traced back, and the resynthesized image is almost identical to the query image. The corresponding absolute error map also contains predominantly black regions, with an average absolute error below 15. In contrast, query images from other sources cannot be correctly traced, resulting in resynthesized images and absolute error maps that exhibit significant noise. This pronounced difference enables reliable determination of whether a query image was generated by the proposed framework using a predefined threshold. Table V shows the mean absolute error (MAE) for different query images. The MAE of images generated by the proposed framework ranges from 2.65 to 13.28, whereas the MAE of other images ranges from approximately 30 to 90. Based on this observation, the threshold is set to 20, which is approximately the midpoint between the maximum MAE of the proposed framework and the minimum MAE of the other images. Fig. 7 further compares the proposed framework with state-of-the-art AI image detectors [23–27, 29], using real images as real samples and images generated by the proposed framework as fake samples. The results show that both the real and fake accuracies of the proposed framework are 100%, outperforming CNNSpot [23], UFD [24], and VIB-Net [27]. These results demonstrate that the proposed framework achieves near state-of-the-art detection performance without relying on additional detectors, using only round-trip consistency verification to protect model ownership.

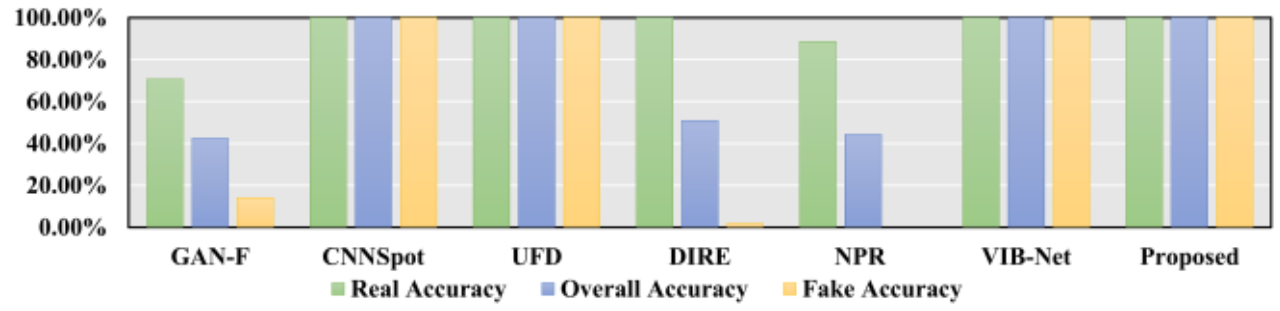

**Fig. 7.** Detection accuracy comparison with state-of-the-art AI image detectors.

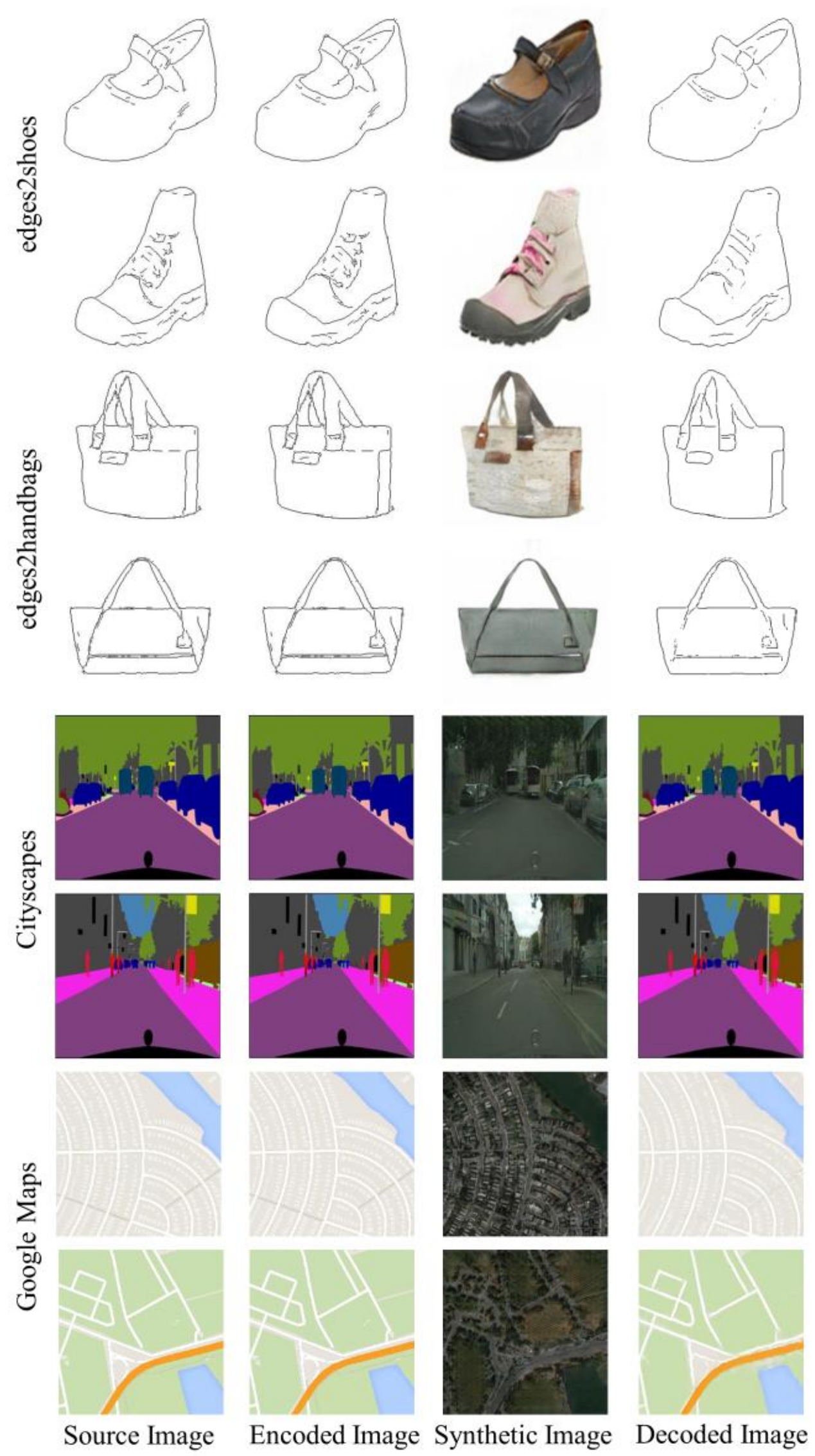

**Fig. 8.** Visualization of the edges2shoes, edges2handbags, Cityscapes, and Google Maps datasets.

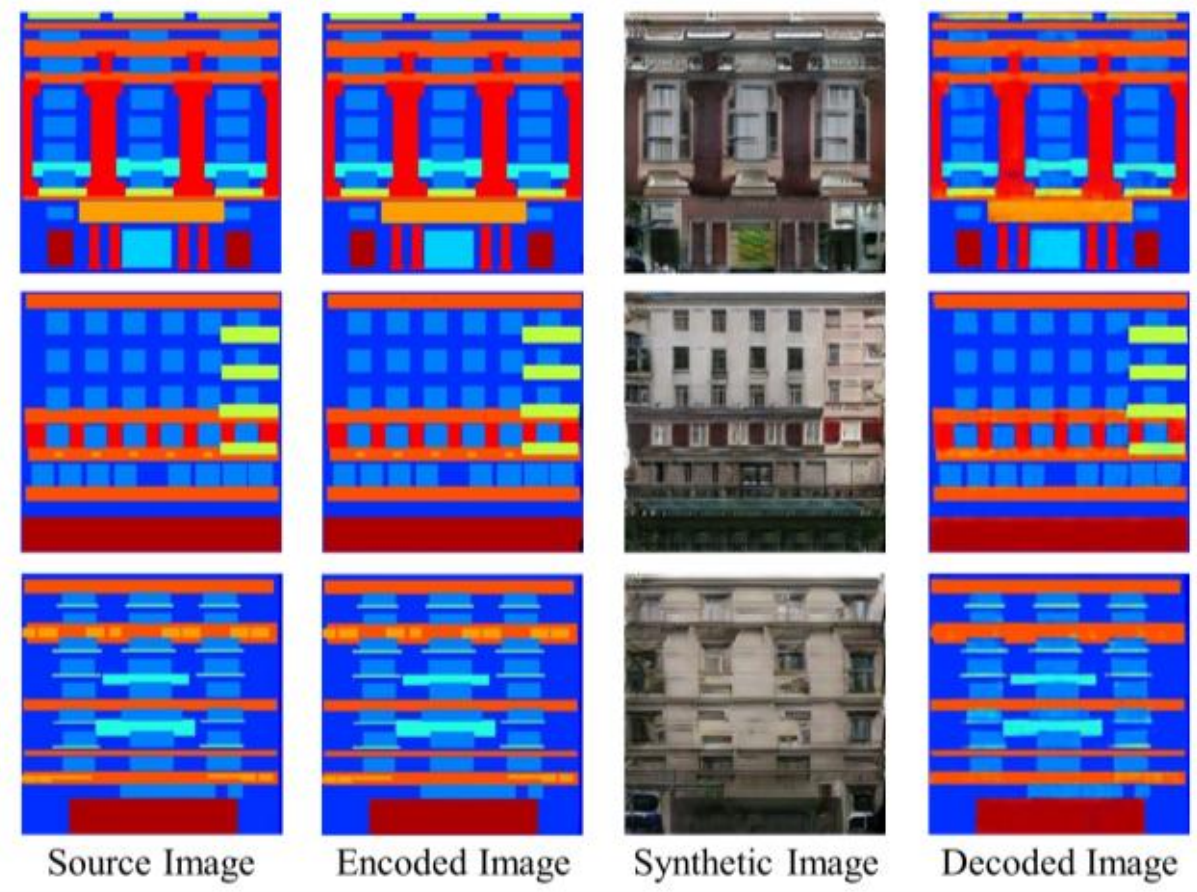

**Fig. 9.** Visualization of the CycleGAN model.

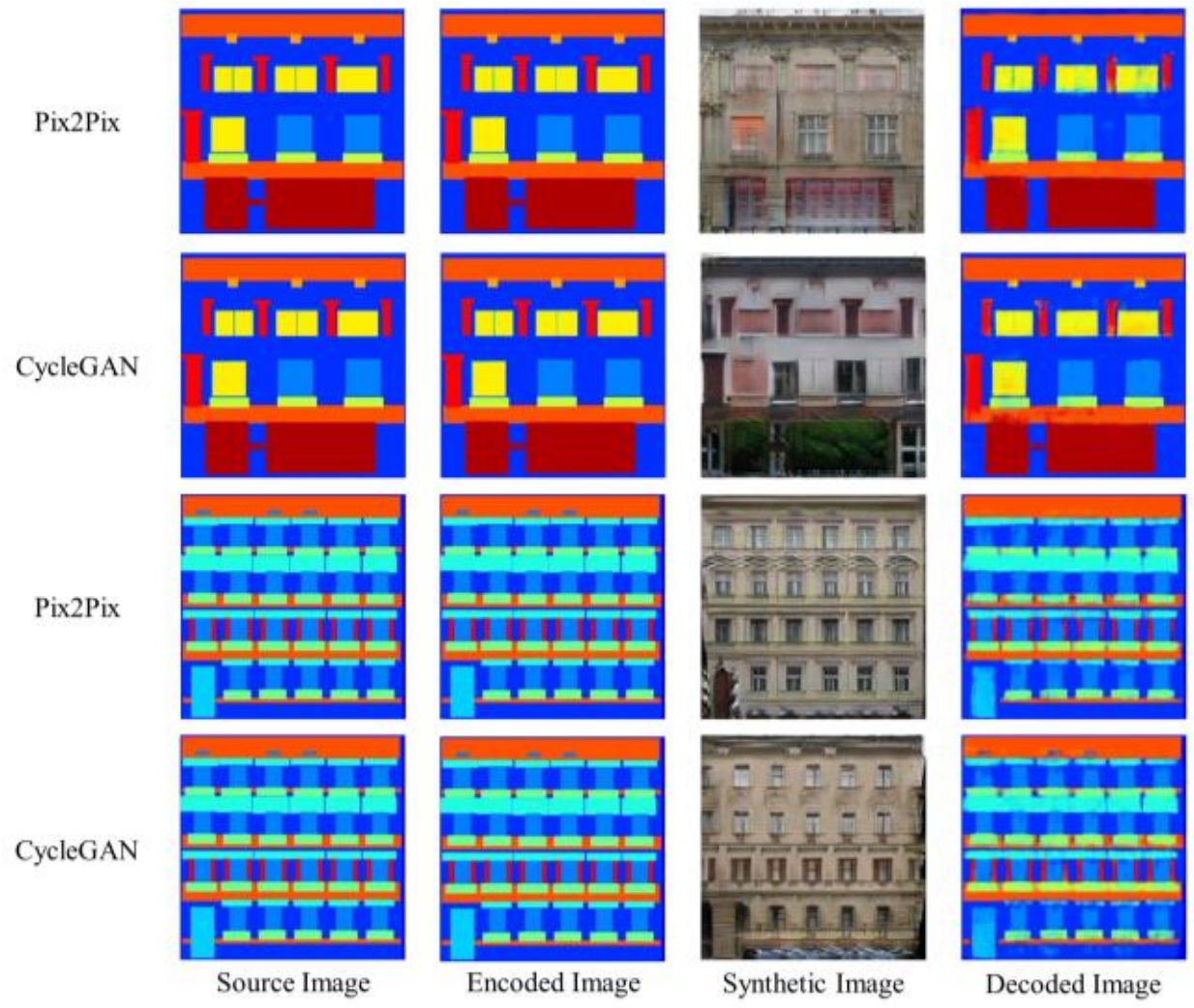

**Fig. 10.** Generalizability evaluation with both Pix2Pix and CycleGAN models.

*E. Generalizability across Generative Models and Tasks*

This subsection examines whether the proposed framework generalizes across different generators and image-to-image translation tasks, beyond the default experimental configuration. To further evaluate its applicability, we conducted experiments using the Pix2Pix [11] model on the edges2shoes [35], edges2handbags [36], Cityscapes [37], and Google Maps [11] datasets, and using the CycleGAN [12] model on the CMP Facade Database [34]. Fig. 8 present the results for Pix2Pix, showing that the decoded images closely trace the source images, while the synthetic images maintain high fidelity.

Fig. 9 presents the results for CycleGAN, which similarly exhibit high fidelity and strong traceable visual quality. To further evaluate generalizability across multiple generators, the proposed framework was trained using both Pix2Pix and CycleGAN as fixed generators. Fig. 10 shows the results, demonstrating that, given the same source image, each generator successfully produces its respective output. Furthermore, the same decoder can process images generated by both generators, yielding decoded images that closely resemble the source. These results indicate that a single encoder–decoder pair is sufficient to handle all generators during training, eliminating the need to train separate encoder–decoder pairs for each generator. This design makes the framework particularly well suited for the MLaaS platform, which supports multiple generators simultaneously.

*F. Comparison with Watermarking Schemes*

To enable a fair and unified comparison with representative watermarking-based provenance verification schemes, we adopt the Accuracy and the Area Under the Receiver Operating Characteristic Curve as metrics for measuring discriminative capability. While all schemes achieve high performance, the interpretation differs according to their detection mechanisms. For watermarking baselines such as HiDDeN [15], StegaStamp [16], and MBRS [17], Accuracy is calculated using a binary classification approach where a sample is classified as positive if the proportion of correctly extracted watermark bits reaches a threshold of 0.85. Additionally, their Area Under the Curve is computed at the bit level using the continuous outputs of the decoder, reflecting the model's ability to distinguish different bit signals. In contrast, the proposed framework evaluates provenance verification rather than bit decoding, using reconstruction residuals derived from a self-referential reconstruction process to assess the distinguishability of the source generator. In both cases, the Area Under the Curve quantifies the separation between positive and negative samples.

Table VI presents a comprehensive quantitative analysis comparing the proposed framework with representative watermarking baselines. The proposed method achieves a perfect Accuracy and Area Under the Curve of 1.00, matching HiDDeN and MBRS and slightly outperforming StegaStamp at 0.98, indicating that residual-based provenance verification is comparable in reliability to bit-error-driven watermark detection. Furthermore, Table VI reports the Peak Signal-to-Noise Ratio at both the input and output stages, which clearly illustrates the fundamental difference in modification stages through the entries marked as not applicable (N/A). Because the baselines embed watermarks into the already generated images, evaluating input fidelity is not applicable to their pipeline. However, their generated outputs incur measurable distortion, resulting in output PSNR values of 38.27 dB for HiDDeN, 29.28 dB for StegaStamp, and 44.80 dB for MBRS. Conversely, the proposed framework is a pre-generation approach that encodes information into the input source, resulting in an initial input PSNR of 39.13 dB. Consequently, the generated output requires no subsequent modification, rendering output PSNR evaluation not applicable.

Table VII provides a qualitative analysis highlighting the methodological advantages of the proposed framework over traditional watermarking baselines. The baselines operate under a post-generation paradigm that inherently requires modifying the generated output images. Because these methods rely solely on extracting embedded metadata from the altered output, they exhibit low verdict explainability and lack a direct reconstructive link to the original generator. Additionally, such techniques are strictly limited to generative model attribution. Conversely, the proposed framework employs a pre-generation methodology based on input modification. This pre-generation approach achieves high verdict explainability by verifying provenance through re-synthesis consistency, which provides a direct reconstructive link to the source generator. Furthermore, unlike the baselines, the proposed framework uniquely enables source image tracing by successfully recovering both the originating generative model and the specific source image responsible for the generated content.

TABLE VI
QUANTITATIVE ANALYSIS OF THE PROPOSED FRAMEWORK WITH WATERMARKING BASELINES

| Scheme | Accuracy | AUC | PSNR (input) | PSNR (output) |
| --- | --- | --- | --- | --- |
| HiDDeN | 1.00 | 1.00 | N/A | 38.27 |
| StegaStamp | 0.99 | 0.98 | N/A | 29.28 |
| MBRS | 1.00 | 1.00 | N/A | 44.80 |
| Proposed | 1.00 | 1.00 | 39.13 | N/A |

TABLE VII
QUALITATIVE ANALYSIS OF THE PROPOSED FRAMEWORK WITH WATERMARKING BASELINES

| Scheme | Generative Model Attribution | Source Image Tracing | Methodology | Verdict Explainability |
| --- | --- | --- | --- | --- |
| HiDDeN | √ | × | Post-generation | Low |
| StegaStamp | √ | × | Post-generation | Low |
| MBRS | √ | × | Post-generation | Low |
| Proposed | √ | √ | Pre-generation | High |

- Post-generation: Output modification.
- Pre-generation: Input modification.
- Low Explainability: Verified by extracted metadata, without a direct visual/reconstructive link to the generator.
- High Explainability: Verified by re-synthesis consistency, providing a direct and reconstructive link to the source generator.

## V. CONCLUSION

This paper presents a self-referential retrosynthesis framework for explainable AI provenance forensics, enabling reliable source tracing and verification of AI-generated content without modifying the original generator. Unlike conventional watermarking, detection, or fingerprinting approaches, the proposed framework integrates a jointly optimized encoder-decoder pair with a fixed, deployed generator to establish a self-embedding mechanism. This design allows client-provided source images to be intrinsically propagated through the generation process and subsequently recovered from the generated outputs. Experimental results demonstrate that the framework produces synthetic images with visual fidelity comparable to the original generator outputs, while retrosynthetic reconstruction accurately traces back to the source inputs. Round-trip consistency verification further provides interpretable evidence of content provenance, achieving near state-of-the-art detection accuracy without additional classifiers. Although validated on Pix2Pix and CycleGAN models across several benchmark datasets, future work will extend the framework to broader generative architectures and application scenarios, enhancing its applicability for explainable generative AI forensics in MLaaS environments.

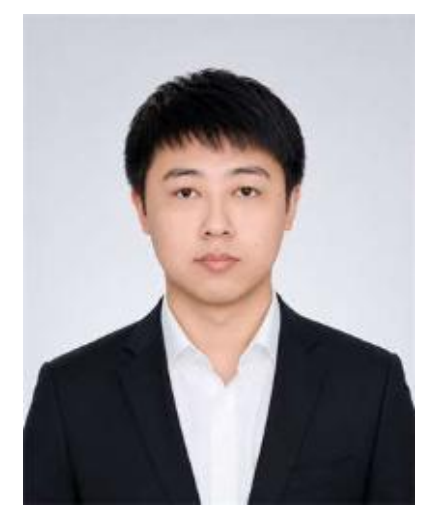

**Yijie Lin** received the B.S. degree in Computer Science and Information Engineering from National Pingtung University, Pingtung, Taiwan, in 2022. He is currently pursuing the Ph.D. degree in Information Engineering and Computer Science at Feng Chia University, Taichung, Taiwan. His research interests include artificial intelligence, steganography, secret image sharing, image processing, information security, and computer vision.

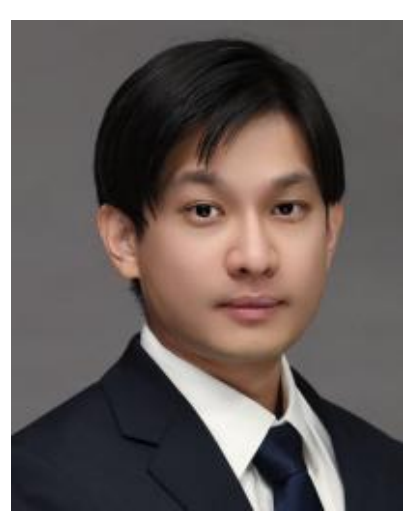

**Ching-Chun Chang** received the PhD in Computer Science from the University of Warwick, UK, in 2019. He is currently affiliated with the National Institute of Informatics, Japan, as a Project Assistant Professor. He also serves as a Visiting Researcher at Peking University, China, and a Distinguished Professor at Hangzhou Dianzi University, China. He is a Senior Member of the IEEE. He participated in a Short-Term Scientific Mission supported by European Cooperation in Science and Technology Actions at the Faculty of Computer Science, Otto von Guericke University of Magdeburg, Germany, in 2016. He was granted the Marie-Curie Fellowship and participated in a Research and Innovation Staff Exchange scheme supported by Marie Skłodowska-Curie Actions at the Department of Electrical and Computer Engineering, New Jersey Institute of Technology, USA, in 2017. He was a Visiting Scholar with the School of Computing and Mathematics, Charles Sturt University, Australia, in 2018, and with the School of Information Technology, Deakin University, Australia, in 2019. He was a Research Fellow with the Department of Electronic Engineering, Tsinghua University, China, in 2020. His research interests include artificial intelligence, biometrics, cryptography, cybersecurity, evolutionary computation, forensics, information theory, steganography, and watermarking.

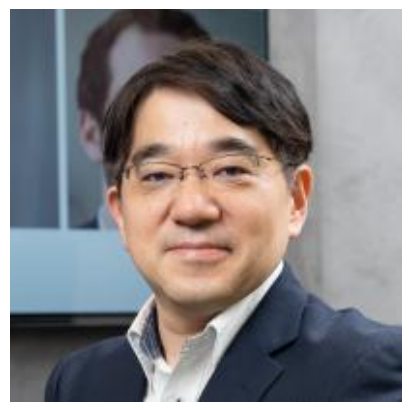

**Isao Echizen** received B.S., M.S., and D.E. degrees from the Tokyo Institute of Technology, Japan, in 1995, 1997, and 2003, respectively. He joined Hitachi, Ltd. in 1997 and until 2007 was a research engineer in the company's systems development laboratory. He is currently a director and a professor of the Information and Society Research Division, the National Institute of Informatics (NII), a director of the Global Research Center for Synthetic Media, the NII, a professor in the Department of Information and Communication Engineering, Graduate School of Information Science and Technology, the University of Tokyo, and a professor in the Graduate Institute for Advanced Studies, the Graduate University for Advanced Studies (SOKENDAI), Japan. He was a visiting professor at the University of Freiburg, Germany, and at the University of Halle-Wittenberg, Germany. He is currently engaged in research on AI security, multimedia security, and multimedia forensics. He is a research director in the CREST FakeMedia project and in the K Program SYNTHETIQ X, Japan Science and Technology Agency (JST). He received the Commendation for Science and Technology by the Minister of Education, Culture, Sports, Science and Technology (Research category) in 2025, the Best Paper Award from the IEICE in 2023, the Best Paper Awards from the IPSJ in 2005 and 2014, the IPSJ Nagao Special Researcher Award in 2011, the DOCOMO Mobile Science Award in 2014, the Information Security Cultural Award in 2016, and the IEEE Workshop on Information Forensics and Security Best Paper Award in 2017. He was a member of the Information Forensics and Security Technical Committee of the IEEE Signal Processing Society. He is the IEICE Fellow, the IPSJ Fellow, the IEEE Senior Member, and the Japanese representative on IFIP and on IFIP TC11 (Security and Privacy Protection in Information Processing Systems), a vice president of APSIPA, and an editorial board member of the *IEEE Transactions on Dependable and Secure Computing*, the *EURASIP Journal on Image and Video Processing*, and the *Journal of Information Security and Applications, Elsevier*.

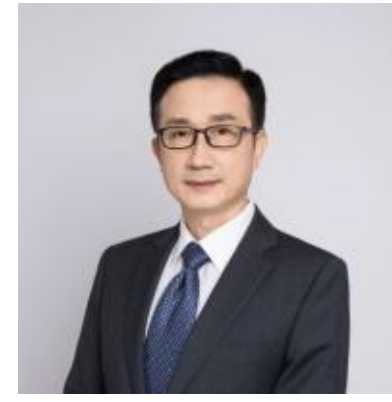

**Hui Li** (Senior Member, IEEE) received the B.Eng. and M.S. degrees from Tsinghua University in 1986 and 1989, respectively, and Ph.D. degree from The Chinese University of Hong Kong in 2000. He is a Professor at Peking University, Fellow of IET. He is also the Double hired Professor of the Peng Cheng Laboratory, the Director of Shenzhen Key Lab of Information theory & Future Internet architecture, the PKU Lab of China Environment for Network Innovations (CENI), National Major Research Infrastructure, Shenzhen Eng. Lab of Converged Networks. His research interests include future network architecture, network protocols, and cyberspace security.

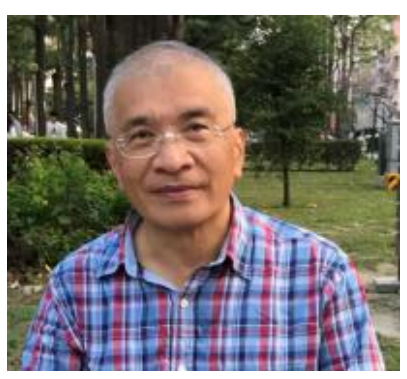

**Chin-Chen Chang** has worked on many different topics in information security, cryptography, multimedia image processing and published several hundreds of papers in international conferences and journals and over 40 books. He was cited over 52,250 times and has an h-factor of 104 according to Google Scholar. Several well-known concepts and algorithms were adopted in textbooks. He also worked with the National Science Council, Ministry of Technology, Ministry of Education, Ministry of Transportation, Ministry of Economic Affairs and other Government agencies on more than 100 projects and holds 40 patents. He served as Honorary Professor, Consulting Professor, Distinguished Professor, Guest Professor at over 60 academic institutions and received Distinguished Alumni Award's from his Alma Master's. He also served as Editor or Chair of several international journals and conferences and had given almost a thousand invited talks at institutions including Chinese Academy of Sciences, Academia Sinica, Tokyo University, Kyoto University, National University of Singapore, Nanyang Technological University, The University of Hong Kong, National Taiwan University and Peking University. Professor Chang has mentored 7 postdoctoral, 72 PhD students and 209 master students, most of whom hold academic positions at major national or international universities. He has been the Editor-in-Chief of Information Education, a magazine that aims at providing educational materials for middle-school teachers in computer science. He is a leader in the field of information security of Taiwan. He founded the Chinese Cryptography and Information Security Association, accelerating information security the application and development and consulting on the government policy. He is also the recipient of several awards, including the Top Citation Award from Pattern Recognition Letters, Outstanding Scholar Award from Journal of Systems and Software, and Ten Outstanding Young Men Award of Taiwan. He was elected as a Fellow of IEEE in 1998, a Fellow of IET in 2000, a Fellow of CS in 2020, an AAIA Fellow in 2021, a Member of the Academy of Europe (AE) in 2022, a Member of the European Academy of Sciences and Arts (EASA) in the same year and a Member of the U.S. National Academy of Artificial Intelligence (NAAI) in 2024 . In 2023, he was awarded the Honorary Ph.D. Degree in Engineering, National Chung Cheng University, Taiwan.